\documentclass[lettersize,journal]{IEEEtran}
\usepackage{amsmath,amsfonts}
\usepackage{algorithmic}
\usepackage{algorithm}
\usepackage{array}
\usepackage{subcaption}
\usepackage[caption=false,font=normalsize,labelfont=sf,textfont=sf]{subfig}
\usepackage{textcomp}
\usepackage{stfloats}
\usepackage{url}
\usepackage{booktabs}
\usepackage{verbatim}
\usepackage{graphicx}
\usepackage{cite}
\begin{document}

\title{A LightGBM-Incorporated Absorbing Boundary Conditions for the Wave-Equation-Based Meshless Method}

\author{Qi-Chang Dong,~\IEEEmembership{Graduate Student Member,~IEEE,} Zhizhang David Chen,~\IEEEmembership{Fellow,~IEEE,} Jun-Feng Wang,~\IEEEmembership{Member,~IEEE,}

\thanks{Manuscript received xxxx; revised xxxx.This article is a preprint going to be submitted to IEEE ANTENNAS AND WIRELESS PROPAGATION LETTERS. All results and methods are solely developed by the authors. Please cite appropriately if referring to this work. It has not yet been peer-reviewed and may contain minor errors or revisions.\textit{(Corresponding author: Zhizhang David Chen)}

Qi-Chang Dong is with deparment of Electrical and Electronic Engineering,the Hong Kong Polytechnic University and also with  Eastern Institute of Technology, Ningbo, 315200, Zhejiang, P. R. China.

Zhizhang David Chen is with the Dalhousie University, Halifax, Canada.
 
Jun-Feng Wang is with the School of Optoelectronic Engineering, Chongqing University of Posts and Telecommunications, Chongqing 400065, China.

 }
}

\markboth{A preprint submitted to IEEE ANTENNAS AND WIRELESS PROPAGATION LETTERS}%
{Shell \MakeLowercase{\textit{et al.}}: A Sample Article Using IEEEtran.cls for IEEE Journals}

\IEEEpubid{xxxxx-\copyright~ IEEE}

\maketitle

\begin{abstract}
A LightGBM-Incorporated absorbing boundary condition (ABC) computation approach for the wave-equation-based the radial point interpolation meshless (RPIM) method  is proposed to simulate wave propagation in open space during the computation process. Different strageties are implemented for replacing the conventional perfectly matched layers (PMLs) in the computational domain. In this work, the model is used to predict the field components on the boundary at each time step to improve computational efficiency. The effectiveness and high efficiency of our method is verified by numerical experiments.

\end{abstract}


\section{Introduction}

Numerical methods, such as the finite-difference time-domain (FDTD) method \cite{FDTD} and finite element method (FEM) \cite{FEM}, have been extensively used for modeling electromagnetic (EM) problems. Among various numerical approaches, node-based meshless methods \cite{meshless1,meshless2,meshless3,Yu001,Yu002} have attracted considerable attention due to their flexibility in node distribution and independence from conventional mesh constraints.

Despite these advantages, computational domains in practical simulations must be restricted to finite sizes due to limited storage capabilities. Consequently, absorbing boundary conditions (ABCs) \cite{ABC1,ABC2} are essential to simulate continuous electromagnetic wave propagation beyond computational boundaries to minimize numerical reflections. Among numerous ABC schemes, perfectly matched layers (PMLs) \cite{PML} are widely recognized as one of the most effective and robust techniques.

In recent decades, various PML variants have been developed, including complex frequency-shifted PMLs (CFS-PMLs) \cite{CPML}, higher-order PMLs \cite{HPML}, uniaxial PMLs (UPMLs) \cite{UPML} and
so on. Typically, finite-thickness PMLs, usually comprising eight to ten layers, are utilized to effectively attenuate EM reflections. However, increasing PML thickness inevitably enlarges the computational domain, leading to significant additional memory usage and computational time.
\begin{figure}[!t]
\begin{center}
\noindent
  \includegraphics[width=0.5\textwidth,trim=3cm 3cm 4cm 3cm,clip]{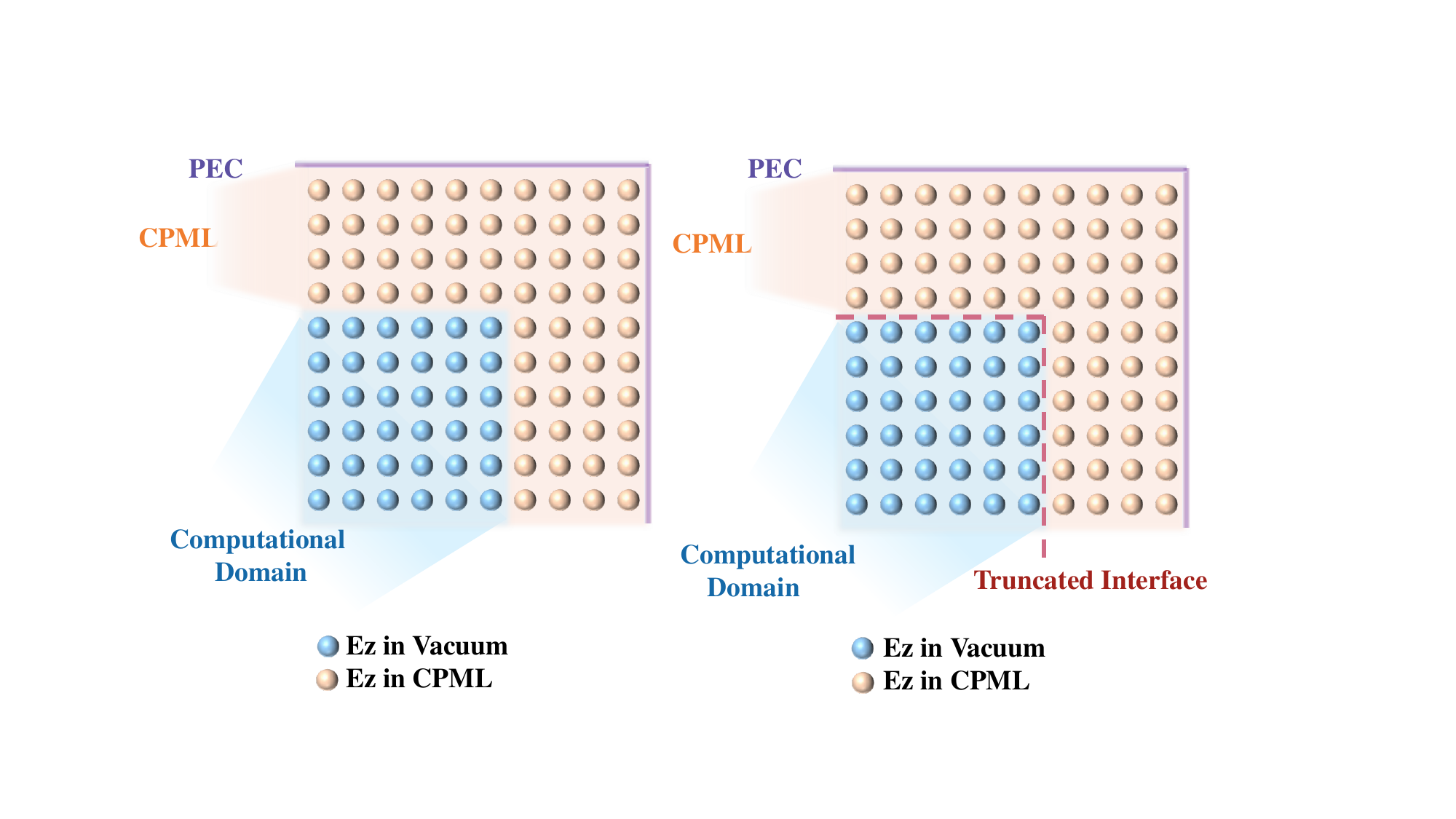}
  \caption{Model example.(a) Original CFS-PML for meshless RPIM(b) Proposed LightGBM-Incorporated ABC:strategy 1}\label{Fig1}
\end{center}
\end{figure}
Recently, rapid advancements in artificial intelligence and machine learning have presented promising solutions for computational electromagnetic challenges \cite{Qi1,Qi2,Limaokun}. According to the universal approximation theorem \cite{NN}, multilayer nonlinear artificial neural networks (ANNs) possess the capability to approximate arbitrary nonlinear, continuous multidimensional functions. Due to their powerful representation and approximation capabilities, ANNs have been extensively applied to solve various complex engineering and scientific problems.

Several neural network-based PML approaches have been proposed for electromagnetic computations \cite{Yao001,Feng001,Feng002,RNN}. Unlike traditional multi-layer PML structures, these neural network-based schemes typically require only a single computational layer, significantly reducing computational complexity, time, and memory consumption without sacrificing accuracy. For instance, a hyperbolic tangent basis function (HTBF)-based ABC model proposed in \cite{Yao001} efficiently predicts boundary component values at each computational step, simplifying the multi-layer PML complexity. Similarly, in \cite{Feng001}, a gradient boosting decision tree-based PMM model has been effectively applied to FDTD methods, addressing low-frequency subsurface sensing problems. Furthermore, a hybrid model combining conventional CFS-PML with recurrent neural networks (RNNs) proposed in \cite{RNN} demonstrated enhanced stability for simulation.

In this letter, one efficient LightGBM-incorporated absorbing boundary condition (ABC) computation method for the wave-equation-based radial point interpolate method is pro-
\newpage
\noindent posed to replace the traditional multi-layers PMLs. Different from previous model in \cite{Feng001,Feng002,RNN,Yao001},  our proposed method uniquely focuses on wave equation-based meshless computations\cite{wangjunfeng003,wangjunfeng002}, only electrical nodes are considered to compute. This innovation results in enhanced computational
efficiency and novelty. The machine learning model, once trained under idealized conditions, effectively simulates wave propagation in unbounded spaces. In this study, three different strategies utilizing the proposed LightGBM-based ABC model are systematically compared with traditional CFS-PML implementations to demonstrate improved performance and efficiency.

\begin{figure}[!t]
\begin{center}
\noindent
  \includegraphics[width=0.5\textwidth,trim=4cm 2cm 3cm 3cm,clip]{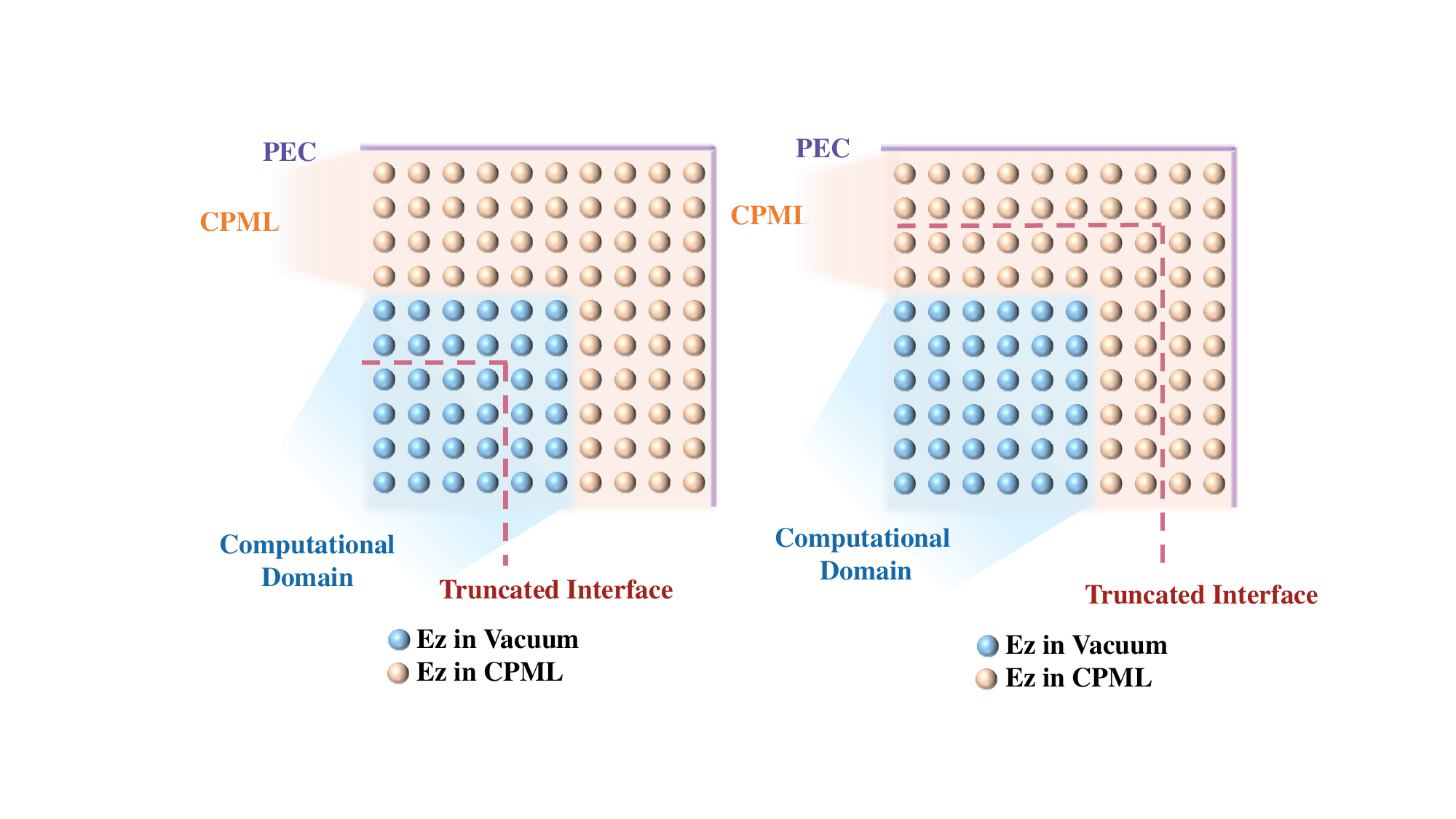}
  \caption{Model example.(a) Proposed LightGBM-incorporated ABC: strategy 2 (b) Proposed LightGBM-incorporated ABC: strategy 3}\label{Fig2}
\end{center}
\end{figure}

\section{methodology}

\subsection{conventional CFS-PML for time-domain meshless method}
 
For convenience, we define electric field nodes in the computational domain like \cite{point-match}. The electric field $E_z$ can be approximated as
\begin{equation}
\begin{aligned}
    E_z(\mathbf{r})=\sum_{m=1}^M r_m(\mathbf{r}) a_m=\mathbf{R}(\mathbf{r}) \mathbf{a}\label{eq13}
\end{aligned}
\end{equation}

where $\mathbf{r}=(x, y)$ is the coordinates of point which $E_z$ is to be interpolated, $r_m(\mathbf{r})$ is the radial basis function associated with node $m$, we choose Gaussian basis function here, $a_m$ is the coefficients to be found, and $M$ represent the number of electric fields nodes within the support domain.

With the meshless formulation \cite{Yu003}, (\ref{eq13}) can be rewritten as
\begin{equation}
E_z^n(\mathbf{r}) =\sum_{m=1}^M \phi_m(x, y) E_{z, m}^n =\boldsymbol{\Phi}(\mathbf{r}) \mathbf{E}_{s z}^n(\mathbf{r})
\label{eq14}
\end{equation}
where $\mathbf{E}_{s z}(\mathbf{r})$ is the unknown electric field value vector at
time step $n$, and the shape function vector is $\Phi(\mathbf{r})=\left[\varphi_1(\mathbf{r}), \varphi_2(\mathbf{r}), \ldots, \varphi_M(\mathbf{r})\right]=$ $\mathbf{R G}^{-1}$ with

\begin{align}
\mathbf{G}=\left[ \begin{array}{ccc}
r_1(x_1, y_1) & \cdots & r_M(x_1, y_1) \\
r_1(x_2, y_2) & \cdots & r_M(x_2, y_2) \\
\vdots & \ddots & \vdots \\
r_1(x_M, y_M) & \cdots & r_M(x_M, y_M)
\end{array} \right] \label{eq15}
\end{align}

The shape function $\Phi(\mathbf{r})$ should be calculated and stored before simulation , the partial derivative can be expressed\cite{ETH} as
\begin{align}
    \frac{\partial \boldsymbol{\Phi}}{\partial x}=\frac{\partial \mathbf{R}}{\partial x} \mathbf{G}^{-1}\label{eq16}
\end{align}

The above meshless method using Gaussian basis function has been proven to be efficient and more accurate than the FDTD method for electromagnetic modeling \cite{Yang001},\cite{Yang002}.

Based on the work in \cite{wangjunfeng001}, the transverse-magnetic case of the 2-D wave equation for electric field $E_Z$ in a homogeneous, lossless, and source-free medium can
be expressed as
\begin{align}
\nabla^2 E_z-\mu \varepsilon \frac{\partial^2 E_z}{\partial t^2}=0\label{eq1}
\end{align}
 we can obtain the equation
\begin{align}
E_{z, i}^{n+1}=2 E_{z, i}^n-E_{z, i}^{n-1}+\frac{\Delta t}{\mu \varepsilon}\left(A_{x 2, i}^{n+\frac{1}{2}}+A_{y 2, i}^{n+\frac{1}{2}}\right)\label{eq:3}
\end{align}

 we can get the final discretization formulations for $\psi_{x 1}$ and $A_{x 1}$.
\vspace{-5pt}
\begin{align}
    \psi_{x 1, i}^{n+\frac{1}{2}} &= c_x \sum_{m=1}^M E_{z, m}^n \frac{\partial \phi_m}{\partial x} + e^{-\alpha_x \Delta t} \psi_{x 1, i}^{n-\frac{1}{2}} \label{eq18a} \\
    A_{x 1, i}^{n+\frac{1}{2}} &= A_{x 1, i}^{n-\frac{1}{2}} + \frac{\Delta t}{\kappa_x} \sum_{m=1}^M E_{z, m}^n \frac{\partial \phi_m}{\partial x} + \Delta t \psi_{x 1, i}^{n+\frac{1}{2}} \label{eq18b}
\end{align}

where $\varphi_m$ represents the shape functions of node $m$ in the local support domain.The formulations for computing auxiliary variables $A_{x 2}, A_{y 1}$, and $A_{y 2}$ can be derived in a similar way, and results are as follows:
\vspace{-5pt}
\begin{align}
    \psi_{x 2, i}^{n+\frac{1}{2}} &= \frac{c_x}{\Delta t} \sum_{m=1}^M \left(A_{x 1, m}^{n+\frac{1}{2}} - A_{x 1, m}^{n-\frac{1}{2}}\right) \frac{\partial \phi_m}{\partial x} + e^{-\alpha_x \Delta t} \psi_{x 2, i}^{n-\frac{1}{2}} \label{eq19}
    \\A_{x 2, i}^{n+\frac{1}{2}} &= A_{x 2, i}^{n-\frac{1}{2}} + \frac{1}{\kappa_x} \sum_{m=1}^M \left(A_{x 1, m}^{n+\frac{1}{2}} - A_{x 1, m}^{n-\frac{1}{2}}\right) \frac{\partial \phi_m}{\partial x} + \Delta t \psi_{x 2, i}^{n+\frac{1}{2}} \label{eq20}\\
    \psi_{y 1, i}^{n+\frac{1}{2}} &= c_y \sum_{m=1}^M E_{z, m}^n \frac{\partial \phi_m}{\partial y} + e^{-\alpha_y \Delta t} \psi_{y 1, i}^{n-\frac{1}{2}} \label{eq21}\\
    A_{y 1, i}^{n+\frac{1}{2}} &= A_{y 1, i}^{n-\frac{1}{2}} + \frac{\Delta t}{\kappa_y} \sum_{m=1}^M E_{z, m}^n \frac{\partial \phi_m}{\partial y} + \Delta t \psi_{y 1, i}^{n+\frac{1}{2}} \label{eq22}\\
    \psi_{y 2, i}^{n+\frac{1}{2}} &= \frac{c_y}{\Delta t} \sum_{m=1}^M \left(A_{y 1, m}^{n+\frac{1}{2}} - A_{y 1, m}^{n-\frac{1}{2}}\right) \frac{\partial \phi_m}{\partial y} + e^{-\alpha_y \Delta t} \psi_{y 2, i}^{n-\frac{1}{2}} \label{eq23}\\
    A_{y 2, i}^{n+\frac{1}{2}} &= A_{y 2, i}^{n-\frac{1}{2}} + \frac{1}{\kappa_y} \sum_{m=1}^M \left(A_{y 1, m}^{n+\frac{1}{2}} - A_{y 1, m}^{n-\frac{1}{2}}\right) \frac{\partial \phi_m}{\partial y} + \Delta t \psi_{y 2, i}^{n+\frac{1}{2}}\label{eq24}
\end{align}

In summary, the computational flowchart of the proposed recursive convolutional CFS-PML computation is as follows:  

1) Calculate the auxiliary variables $\psi_{x1}$ and $\psi_{y1}$ according to equations (7) and (11).

2) Update the auxiliary variables $A_{x1}$ and $A_{y1}$ based on (8) and (12).

3) Compute the auxiliary variables $\psi_{x2}$ and $\psi_{y2}$ using (9) and (13).

4) Update $A_{x2}$ and $A_{y2}$ following (10) and (14).

5) Update the electric field component $E_{z}$ according to (15).

6) Repeat the above steps for subsequent time steps.

Within the CFS-PML medium, the PML parameters are selected and scaled as follows:  
\begin{align}
\sigma_{w,i}\left(r_{w,i}\right) &= \sigma_{w}^{\max}\left(\frac{r_{w,i}}{d}\right)^{n} \label{eq31} \\
\kappa_{w,i}\left(r_{w,i}\right) &= 1+(\kappa_w^{\max}-1)\left(\frac{r_{w,i}}{d}\right)^n \label{eq32} \\
a_{w,i}\left(r_{w,i}\right) &= a_w^{\max}\left(\frac{r_{w,i}}{d}\right) \label{eq33}
\end{align}

where $i$ represents the $i_{th}$ layer of the PML, $w = x$ or $y$, $r_{w,i}$ is the distance between the $i_th$ PML node and the interface between the PML medium and the computation domain, $d$ is the thickness of the CFS-PML region, and $n$ is the order. The choice of $\sigma_{opt}$ \cite{FDTD,CPML} is taken to be $(n+1)(150\pi\Delta s)$ where $s$ is the nodal spacing. The CFS-PML parameters in this paper are $n$ = 4, $\sigma_{max}$ = 2$\sigma_{opt}$, $k_{opt}$ = 5, and $a_{max}$ = 0.05, respectively.

\subsection{the construction of lightGBM based absorbing boundary conditions for the Wave-Equation-Based Meshless Method}
Traditional PML is often used to truncate computational domains, but often requires multiple layers such as 8 or 10 and iterative computation of intermediate variables during computation like previous $\psi_{x1}$,$\psi_{y1}$,$A_{x1}$,$A_{y1}$,$\psi_{x2}$,$\psi_{y2}$,$A_{x2}$ and $A_{y2}$, which raises the complexity of the computation. In practice, the parameters need to be adjusted according to the scenario to get a better result, which is very costly and time-consuming, so here we introduce neural networks, whose approximation performance has been proved in \cite{NN}, to replace the multilayer PML to improve the efficiency of the computation.

Here, we have chosen to use the wave-equation corresponding to the maxwell system of equations, in which only the electric field and not the magnetic field values need to be computed.

For example, we trained two different lightGBM model for the corner cell and boundary cell, the input and output of the right boundary cell in Figure1 is 
\begin{equation}
\left\{
\begin{aligned}
x &= \Big[E_z\big|_{i,j-1}^{s-1},E_z\big|_{i+1,j-1}^{s-1},E_z\big|_{i-1,j-1}^{s-1} , E_z\big|_{i,j-1}^s, E_z\big|_{i+1,j-1}^s\\
   &\quad, E_z\big|_{i-1,j-1}^s, E_z\big|_{i,j}^{s-1},E_z\big|_{i+1,j}^{s-1},E_z\big|_{i-1,j}^{s-1} , E_z\big|_{i,j}^s, E_z\big|_{i+1,j}^s\\
   &\quad, E_z\big|_{i-1,j}^s\Big] 
   \\
y &= \Big[ E_z\big|_{i,j+1}^{s+1} \Big]
\label{eq34}
\end{aligned}
\right.
\end{equation}

Here the $s$ represent the $s-th$ time step in the simulation process. And $i$ and $j$ is the index of $X$ and $Y$ coordinates, indicating $x_{i}$ and $y_{j}$, respectively.

The input and output of the right corner cell in Figure1 is 
\begin{equation}
\left\{
\begin{aligned}
&x = \Big[E_z\big|_{i+1,j-1}^{s-1},E_z\big|_{i+2,j-1}^{s-1},E_z\big|_{i+1,j-2}^{s-1} , E_z\big|_{i+1,j-1}^s\\&\quad, E_z\big|_{i+2,j-1}^s
   ,E_z\big|_{i+1,j-2}^s, E_z\big|_{i,j}^{s-1},E_z\big|_{i+1,j}^{s-1},E_z\big|_{i,j-1}^{s-1} \\&\quad, E_z\big|_{i,j}^s, E_z\big|_{i+1,j}^s
   , E_z\big|_{i,j-1}^s\Big] 
   \\
y &= \Big[ E_z\big|_{i-1,j+1}^{s+1}, E_z\big|_{i-1,j}^{s+1}, E_z\big|_{i,j+1}^{s+1} \Big]\label{eq35}
\end{aligned}
\right.
\end{equation}

And in this part, three different truncation strategies can be considered, strategies 1 is truncated at the interface of computational domain and CFS-PML domain in Figure \ref{Fig1}(b),which has been discussed extensively in \cite{Yao001,Feng001,Feng002}. The strategies 2 is an hybrid scheme in Figure \ref{Fig2}(b) proposed by paper \cite{RNN}, which combines the advantage of
conventional method and machine learning, has a better performance even in the late time steps. And the strategies 3 is truncated interface in the inner space in ideal space in Figure \ref{Fig2}(a), which provides a new way to get training dataset even without the formula of traditonal PMLs.

It is worth noting that the data for strategies 1 and 2 can be collected on a traditional computational domain having 20 layers of CFS-PML\cite{Yao001}, which means we need extra cost to obtain the dataset, while the data for strategy 3 can be collected on a sufficiently large and desirable domain, so that we can easily obtain sufficiently accurate dataset without implementing any traditional PML or any other absorbing boundary conditions.

Gradient Boosting Decision Trees (GBDT) \cite{GBDT} is a widely adopted tree-based machine learning algorithm. Nonetheless, their performance and scalability are often limited when applied to large-scale datasets. This limitation stems from the need to scan all data instances for each feature to calculate information gain at every possible split, a process that is highly time-consuming.
Then, different strategies, including histogram-based algorithm, gradient-based one-side sampling, greedy bundling, and merge exclusive features, were proposed in LightGBM\cite{LightGBM}. It is reported that lightGBM can accelerate the training process by up to over 20 times than GBDT while achieving almost the same accuracy\cite{LightGBM}. Due to space limitations, only histogram-based algorithm is introduced here. For a comprehensive understanding, readers are referred to the original paper\cite{LightGBM}.

\begin{figure}
\begin{center}
\noindent
  \includegraphics[width=0.48\textwidth]{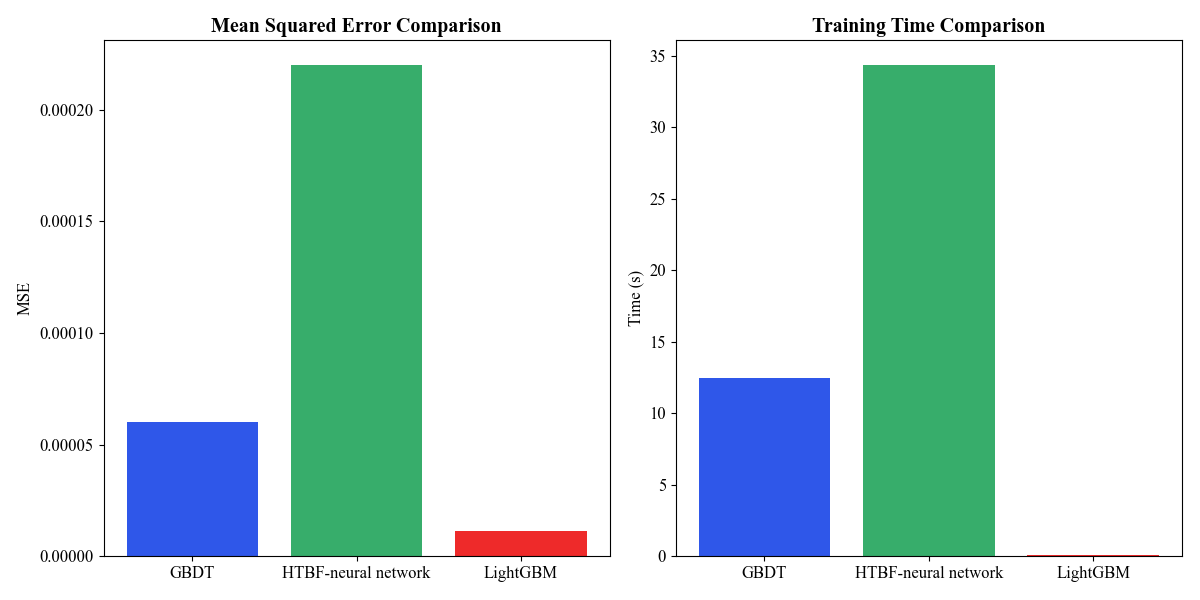}
  \caption{MSE and training time comparison between different models}\label{Fig3}
\end{center}
\end{figure}

\begin{figure}
\begin{center}
\noindent
  \includegraphics[width=0.5\textwidth]{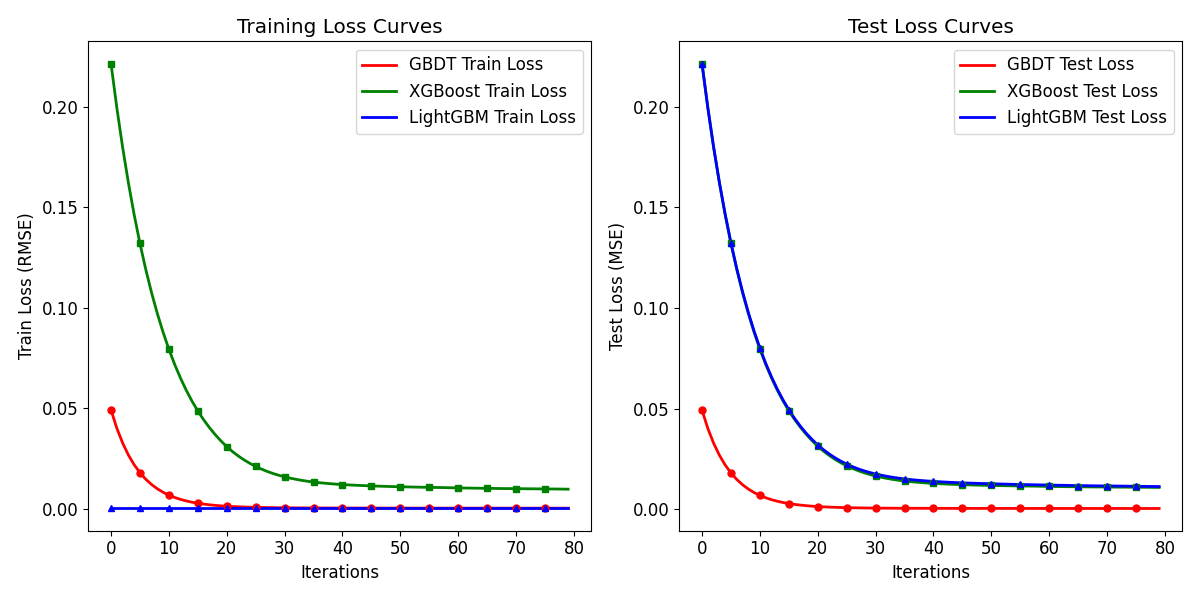}
  \caption{Relative error of Ez at each epoch in the training process.}\label{Fig4}
\end{center}
\end{figure}

At each node, GBDT needs to compute the loss of all samples and find the optimal split point $s^*$:
\begin{align}
   s^*=\arg\min_s\sum_{x_i\in D_{\mathrm{left}}}(y_i-\bar{y}_{left})^2+\sum_{x_i\in D_{right}}(y_i-\bar{y}_{right})^2\label{eq35} 
\end{align}

LightGBM uses histogram-based optimization to speed up feature splitting. Instead of scanning all unique feature values, it groups continuous values into discrete bins, reducing computation from \( O(n) \) to \( O(k) \).

For a feature \( j \), a histogram with \( k \) bins is built:
\begin{align}
  H_j(b) &= \sum_{i=1}^{n} h(x_{ij} \in B_b) \label{eq36}  
\end{align}

where \( H_j(b) \) represents the number of samples in bin \( b \). \( B_b \) is the range of values in bin \( b \). The indicator function:
\begin{align}
h(x_{ij} \in B_b) &=
\begin{cases}
1, & \text{if } x_{ij} \in B_b \\
0, & \text{otherwise}
\end{cases} \label{eq37}
\end{align}

For each bin \( b \), LightGBM accumulates gradient \( G_b \) and Hessian \( H_b \):
\begin{align}
G_b = \sum_{x_i \in B_b} g_i, \quad H_b = \sum_{x_i \in B_b} &h_i \label{eq38}\\
 g_i = \frac{\partial L(y_i, F(x_i))}{\partial F(x_i)} \label{eq39}\\
 h_i = \frac{\partial^2 L(y_i, F(x_i))}{\partial F(x_i)^2} 
\end{align}
 
Here $g_i$ represent the gradient. $h_i$ is the Hessian.

LightGBM selects the best split point \( s^* \) by minimizing:
\begin{align}
s^* = \arg\min_s \left[ \frac{(G_{left})^2}{H_{\text{left}} + \lambda} + \frac{(G_{right})^2}{H_{right} + \lambda} - \frac{G_{total}^2}{H_{total} + \lambda} \right] \label{eq39}
\end{align}

 \( G_{left}, G_{right} \) are gradient sums for left and right child nodes. \( H_{left}, H_{right} \) are Hessian sums. \( \lambda \) is a regularization parameter.

\begin{algorithm}[t]
\caption{LightGBM-based ABC Model for the Wave-Equation-Based Meshless Method}\label{alg:alg1}
\begin{algorithmic}
\STATE 
\STATE {\textbf{TRAINING PROCESS}}
\STATE \textbf{Input:} Training dataset \( D = \{(x_i, y_i)\}_{i=1}^{n} \), number of trees \( M \), max depth \( d \), learning rate \( \eta \)

\STATE \textbf{Initialize} model with constant value: \( F_0(x) = \arg\min_c \sum_{i=1}^{n} L(y_i, c) \),for $m = 1$ to $M$
    \STATE \textbf{Loop} Compute first-order gradient$g_i$ and second-order Hessian $h_i$
    \STATE Construct histogram bins for each feature \( j \):
    \[H_j(b) = \sum_{i=1}^{n} 1(x_{ij} \in B_b)\]
    \STATE Accumulate gradient and Hessian in each bin:
    \[
    G_b = \sum_{x_i \in B_b} g_i, \quad H_b = \sum_{x_i \in B_b} h_i
    \]
    \STATE Find the best split \( s^* \) by minimizing:
    \[
    s^* = \arg\min_s \left[ \frac{(G_{\text{left}})^2}{H_{\text{left}} + \lambda} + \frac{(G_{\text{right}})^2}{H_{\text{right}} + \lambda} - \frac{G_{\text{total}}^2}{H_{\text{total}} + \lambda} \right]
    \]
    \STATE Grow a new leaf-wise tree \( h_m(x) \) using leaf-wise splitting strategy
    \STATE Update model:
    \[
    F_m(x) = F_{m-1}(x) + \eta h_m(x)
    \]
\STATE \textbf{Return} final model \( F_M(x) \)
\STATE {\textbf{PREDICTING PROCESS}}
\STATE \textbf{Loop}
\STATE predict $E_z$ value at every boundary cell and corner cell with trained model
\STATE Update $E_z$ in computational domain by the Meshless method

\end{algorithmic}
\label{alg1}
\end{algorithm}

\begin{figure}
\begin{center}
\noindent
  \includegraphics[width=0.45\textwidth]{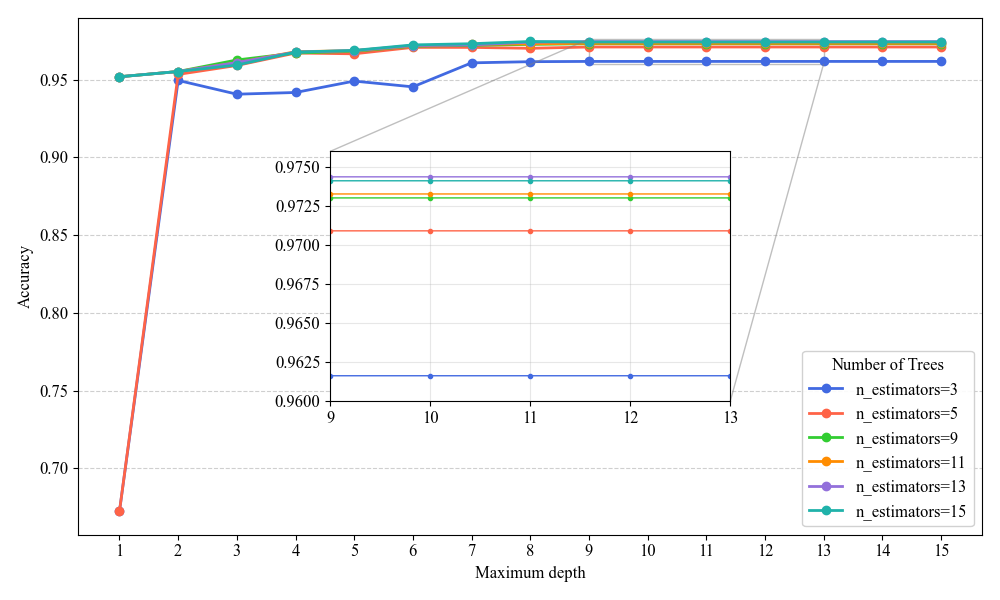}
  \caption{Accuracy of the maximum depth of tree in the training process.}\label{Fig5}
\end{center}
\end{figure}
\begin{figure}
\begin{center}
\noindent
  \includegraphics[width=0.4\textwidth]{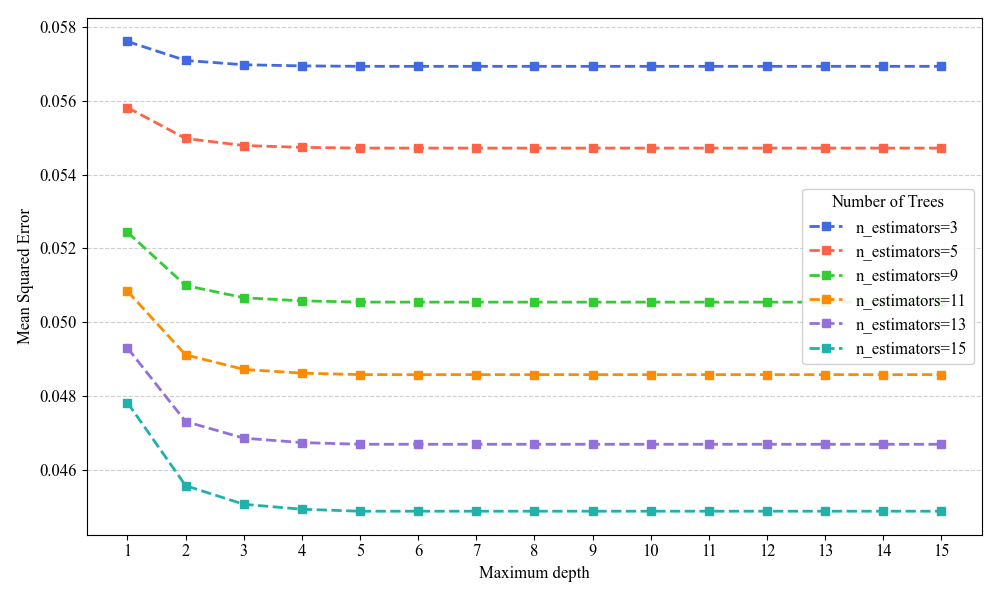}
  \caption{The mean square error of Ez at each epoch in the training process.}\label{Fig6}
\end{center}
\end{figure}

The mean square error is commonly used to evaluate regression tasks and it is defined as
\begin{align}
MSE &= \frac{1}{n} \sum_{i=1}^{n} (y_i - \hat{y}_i)^2 \label{eq41}
\end{align}
The accuracy is measured by the relative error between prediction and true value as follows:
\begin{align}
Accuracy &=1.0 - \frac{\sqrt{\sum_{i=1}^n\left(y_i^2 - \hat{y}_i^2\right)}}{\sqrt{\sum_{i=1}^n y_i^2}} \label{eq:accuracy}
\end{align}

where $y_i$ is the true values of Ez , $\hat{y}_i$ is the prediction of Ez. Besides, $n$ is the size of testing samples.

The results of training process of \cite{Feng001}, \cite{Yao001},and proposed model are compared in Fig.\ref{Fig3}, whether the MSE or the training time of LightGBM is more promising than others. We also went through a series of experiments to determine the hyperparameters of the model such as
the number of estimators. As shown in Fig.\ref{Fig5} and Fig.\ref{Fig6}, the experimental results indicate that 
more estimators can better learn the mapping relationship in the original dataset, but too many estimators 
can cause overfitting and an increase in calculations. Therefore, we chose estimators number of 13 to ensure that the model has a stronger generalisation ability.

 

\section{NUMERICAL RESULTS AND DISCUSSION}
To validate the performance of the proposed scheme for the wave-equation-based meshless method, two TM-polarized current source propagation numerical experiment is conducted.
The first case is to validate the performance of our proposed model with different strageties, and the second case is aimed to apply the trained model to a new scattering problem to validate the generalisation of our model while maintaining accuracy and efficiency.
\subsection{Case 1}
\begin{figure}
\begin{center}
\noindent
  \includegraphics[width=0.4\textwidth]{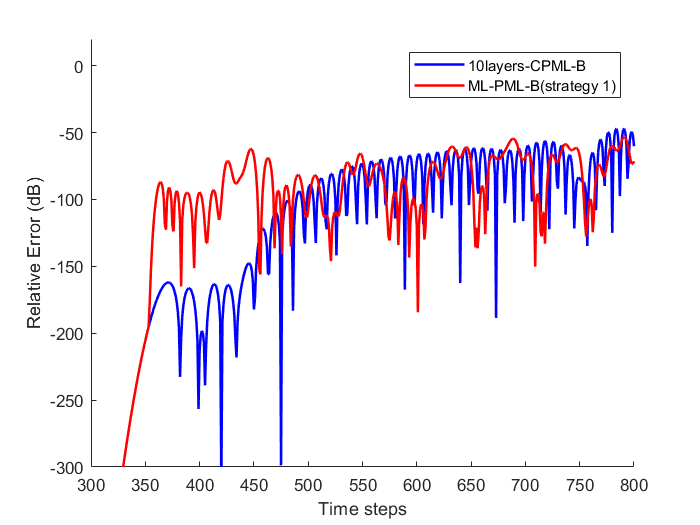}
  \caption{Relative reflection errors among ten-cell conventional CFS-PML and proposed method of test point B.}\label{Fig7}
\end{center}
\end{figure}
\begin{figure}
\begin{center}
\noindent
  \includegraphics[width=0.4\textwidth]{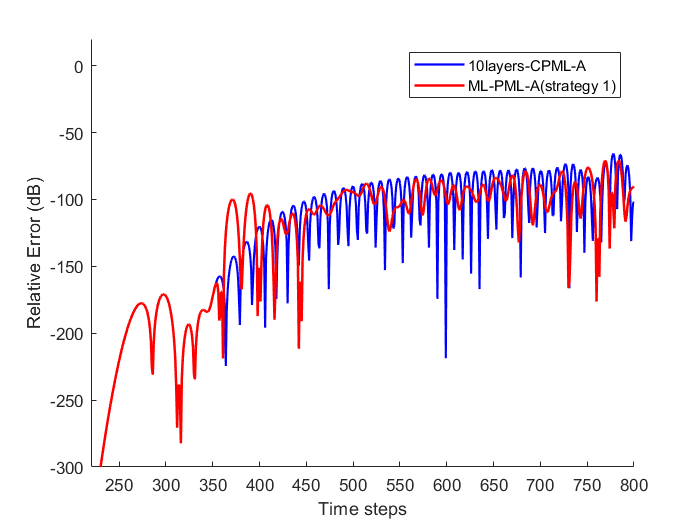}
  \caption{Relative reflection errors among ten-cell conventional CFS-PML and proposed method of test point A.}\label{Fig8}
\end{center}
\end{figure}

Without loss of generality,a wave propagation problem in an unbounded domain is considered, seen in\cite{wangjunfeng001}.
The solution domain of air is discretized with 101 × 101 uniform nodes, and the nodal spacing is $\Delta x=\Delta y=1.5$ mm, and then t step is calculated as $\Delta t=1$ ps. A sinusoidal pulse excited at the center of the solution domain, which can be expressed as $J_z = \sin(2\pi f_mt)$ with $f_m = 10$ GHz. 

The simulated electric fields of 800 time steps are recorded at two test points within the solution domain. Test point A is located at the position one node away from the right PML; test point B is located in the top right corner of the solution domain. To evaluate the performance of the proposed method, errors are defined as
\begin{equation}
    e(t)_{\mathrm{dB}}=20\times\log_{10}\left[\left|E_{z}(t)-E_{z}^{ref}(t)\right|/\left|E_{z}^{ref,\max}\right|\right]
\end{equation}

where $E_z$ represent the recorded electric field component, $E_{z}^{ref}$is the reference field without any reflection effect in a big enough field, and $E_{z}^{ref,\max}$ is the maximum amplitude of $E_{z}^{ref}$.

As shown in Fig.~\ref{Fig7} and Fig.~\ref{Fig8}, the relative reflection error of time-dependent electric field $E_z$ at points A and B is shown when ten layers of conventional CFS-PML absorbing boundary conditions and the proposed model with Strategy 1 are applied. The maximum relative reflection errors at test point B for the ten-cell CFS-PML and the LightGBM-incorporated model are $-47.16$~dB and $-53.32$~dB, respectively. Similarly, at test point A, the maximum relative reflection errors are $-65.56$~dB and $-70.85$~dB, respectively. The generally lower reflection errors observed at test point A can be attributed to its greater distance from the corner of the computational domain, thereby reducing boundary-related reflections. These results demonstrate that the proposed method achieves better absorption performance with lower reflection errors. Furthermore, compared to the conventional method, our model requires less CPU time, as it only uses a single layer to truncate the computational domain.

\vspace{-2mm}
\subsection{Case 2}

\begin{table}
\caption{Computational time (s) for different models\label{tab:table1}}
\centering
\begin{tabular}{
    >{\centering\arraybackslash}p{2.5cm}
    >{\centering\arraybackslash}p{2.5cm}
    >{\centering\arraybackslash}p{2.5cm}
}
\toprule
Methods & Structure & CPU time (s) \\
\midrule
CFS-PML     & 10 layers & 36.98 \\
CFS-PML     & 30 layers & 46.27 \\
Proposed & 1 layer   & 29.36 \\
\bottomrule
\end{tabular}
\end{table}

\begin{figure}[!t]
\begin{center}
\noindent
\includegraphics[width=0.5\textwidth,trim=6cm 3cm 9cm 1cm,clip]{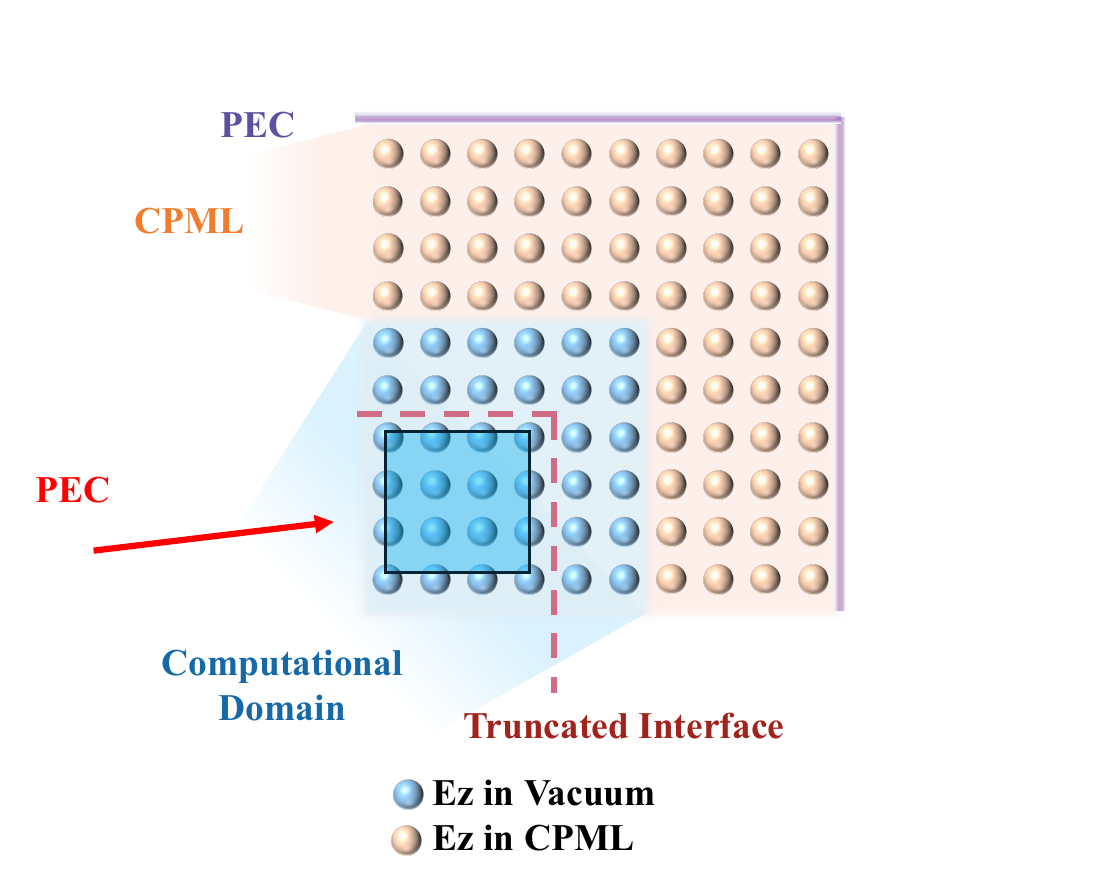}
  \caption{Proposed ANN-Incorporated ABC}\label{Fig9}
\end{center}
\end{figure}

\begin{figure}
  \centering
  \begin{subfigure}{0.5\linewidth}
    \includegraphics[width=\linewidth]{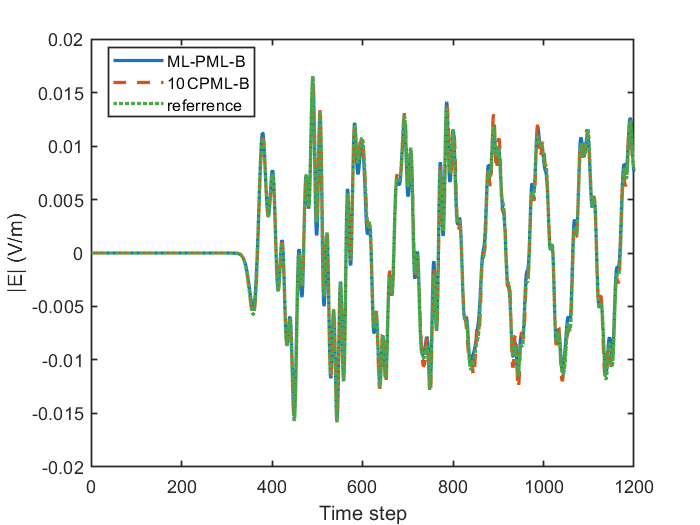}
    \label{Fig11a}
  \end{subfigure}\hfill
  \begin{subfigure}{0.5\linewidth}
    \includegraphics[width=\linewidth]{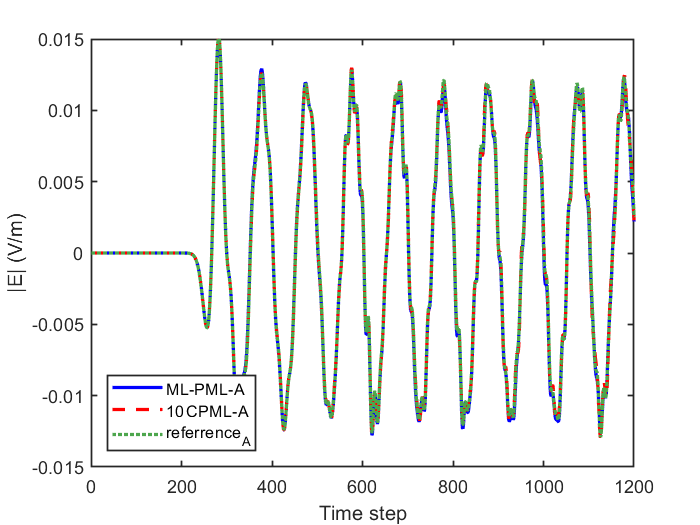}
    \label{Fig11b}
  \end{subfigure}
  \caption{Waveforms at test point A and B}\label{Fig11}
\end{figure}

\begin{figure}[h]
  \centering
  \begin{subfigure}{0.7\linewidth}
    \includegraphics[width=\linewidth]{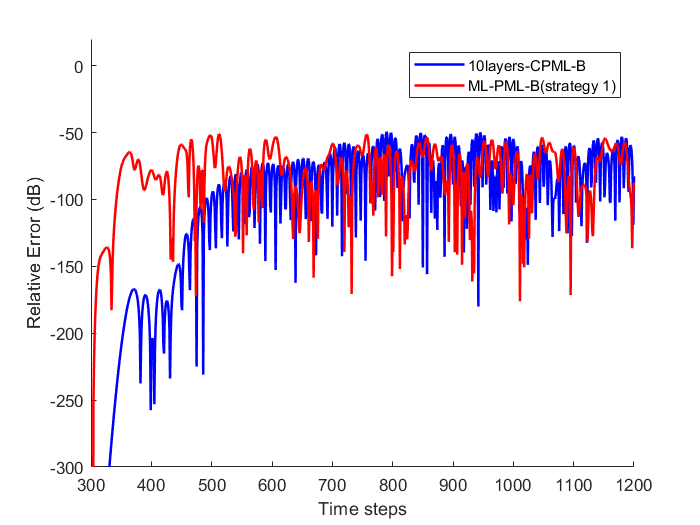}
  \end{subfigure}\hfill
  \begin{subfigure}{0.7\linewidth}
    \includegraphics[width=\linewidth]{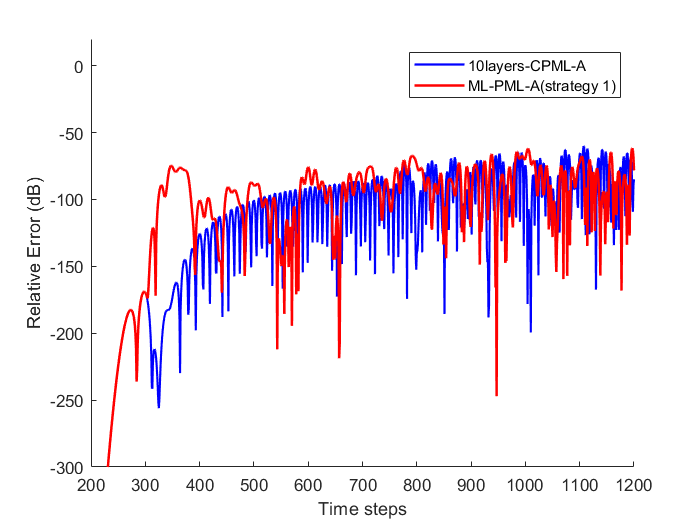}
  \end{subfigure}
  \caption{Relative reflection errors among ten-cell conventional CFS-PML and proposed method of two test points}\label{Fig12}
\end{figure}

The second example is a scattering problem involving a single PEC scatterer with dimensions of $20 \times 20$ mm, located within the same computational domain. The waveforms at test points A and B, observed over 1200 time steps, are presented in Fig.~\ref{Fig11}. As shown, the waveforms produced by the proposed method with strategy 1 are in good agreement with those obtained using the conventional 10-layer CFS-PML. A comparison of the relative reflection errors between the proposed model and the conventional ten-cell CFS-PMLs is illustrated in Fig.~\ref{Fig12}.

The maximum relative reflection errors of the conventional ten-cell CFS-PML and the proposed LightGBM-based PML are $-49.33$~dB and $-51.81$~dB at test point B, and $-60.31$~dB and $-61.89$~dB at test point A, respectively. These results indicate that the proposed LightGBM-incorporated model achieves slightly better absorbing performance compared to the ten-cell conventional CFS-PML. Moreover, since the proposed method requires only one layer to truncate the computational domain, it significantly reduces CPU time consumption, as shown in Table~\ref{tab:table1}. In addition, the efficiency is further improved because the trained model can be reused during the computation process, which greatly reduces the overall computational resource usage.

All the models are implemented on a Intel(R) Core(TM) i7-10750H CPU, time consumption may vary on different devices.

\section{Conclusion}
In this letter, the LightGBM-incorporated CFS-PML method with three different truncation strategies applicable to wave-equation-based meshless method is proposed to improve the efficiency of the unbounded electromagnetic problems. The tree model-based LightGBM that we adopt takes less time and is more accurate in training compared to other machine learning models. The computational domain is well reduced due to the internal truncation of CFS-PML. Therefore, the proposed method not only achieves higher accuracy and efficiency with fewer layers.

\bibliographystyle{IEEEtran}  
\bibliography{cite}

\end{document}